# Geometry-induced Exceptional Point Detached from Fermi Arcs

Yuancheng Zhao[*,1], Jia-Xin Zhong[*,2,§], Jing Lin[1,#], Yun Jing[2,‡], and Kun Ding[1,†]

[1] *Department of Physics, State Key Laboratory of Surface Physics, and Key Laboratory of Micro and Nano Photonic Structures (Ministry of Education), Fudan University, Shanghai 200438, China*

[2] *Graduate Program in Acoustics, The Pennsylvania State University, University Park, PA 16802, USA*

*These authors contributed equally.

§E-mail: Jiaxin.Zhong@psu.edu, #E-mail: jing_lin@fudan.edu.cn

‡E-mail: yqj5201@psu.edu, †E-mail: kunding@fudan.edu.cn

**Abstract**

Exceptional points (EPs), ubiquitous non-Hermitian degeneracies, are central features in band structures where non-Hermitian Fermi arcs connect EPs and eigenvalue knots encircle them. Under open boundary conditions (OBCs), non-Hermitian skin effects enforce complex momenta and yield non-Bloch band structures, introducing EPs unique to OBCs whose origins depend on boundary-driven mechanisms. Here, we reveal both theoretically and experimentally that geometry itself can induce such non-Bloch EPs in a reciprocal non-Hermitian Lieb lattice supporting geometry-dependent skin effects. By analyzing non-Bloch band structures, we find that geometry-induced EPs correspond to saddle points rather than branch points. Branch points, even while carrying OBC eigenenergies, do not yield EPs but manifest as Whitney cusps, a characteristic type of geometric singularity, and Fermi arcs connecting them remain crucial in determining eigenvalue knots. Our measurements of these knots confirm that geometry-induced EPs are detached from the branch points of Fermi arcs, contrasting with their unified counterparts in Bloch systems. Our results establish geometry as an additional degree of freedom for engineering EP-based devices and reveal its fundamental role in shaping non-Bloch band structures.

***Introduction***—Exceptional points (EPs) are spectral singularities unique to non-Hermitian systems, where both eigenvalues and eigenvectors coalesce [1-4]. Beyond Hermitian degeneracies, EPs exhibit nontrivial eigenvalue topology [5-18], leading to many unconventional physical effects [19-24]. Their realization across diverse platforms has stimulated broad interest [25-33], with applications ranging from ultrasensitive sensing [34-39] and controllable lasing [40-42] to robust topological devices [43,44]. In lattice systems, extensive studies investigate EPs under periodic boundary conditions (PBCs), which are branch points of the complex-eigenvalue manifold in the Brillouin zone (BZ) and are termed Bloch EPs [2,3,45]. Due to the no-go theorem, EPs always emerge in pairs connected by branch cuts known as non-Hermitian Fermi arcs (FAs) [46-48]. These EPs and non-Hermitian FAs fundamentally govern the braiding topology of complex energy bands [7,8,10-12].

Under open boundary conditions (OBCs), non-Hermitian skin effect drastically reshapes the spectrum [49-61], erasing Bloch EPs from the OBC spectrum. The non-Bloch band theory, which replaces the BZ with the generalized Brillouin zone (GBZ) [49,62], captures the bulk behavior in this regime and reveals unprecedented topological structures, including the braiding topology of OBC eigenvalue knots [63,64]. EPs may emerge solely in the OBC spectrum as non-Bloch EPs [33,65-67], often assumed to coincide with branch points of Riemann surfaces viewed from the GBZ [68-70], whose branch cuts determine eigenvalue knots [11,19]. However, a recent study shows that the parity-time symmetry transition point under OBCs, if treated as an EP, can be identified from the GBZ geometry and even occurs in single-band cases [71], indicating that a class of non-Bloch EPs is not the branch point of Riemann surfaces. This finding challenges the exceptional geometry arising from EPs and eigenvalue knots that follow, motivating the pursuit of means unique in open boundaries to harness non-Bloch EPs and clarify the related eigenvalue knots.

Here, we theoretically and experimentally demonstrate that, in higher-dimensional lattices, geometry itself induces non-Bloch EPs that are not located at branch points of Riemann surfaces. Nonetheless, their associated branch cuts still govern the formation of OBC eigenvalue knots. Using a reciprocal Lieb lattice supporting geometry-dependent skin effect and Bloch EPs, we find that geometric shapes trigger non-Bloch EPs at energies distinct from Bloch EPs. Unlike Bloch EPs, intersections between GBZs and branch points do not yield non-Bloch EPs but instead form Whitney cusps, a characteristic type of geometric singularity [72-74], in the non-Bloch band structure. These cusps are intrinsically tied to branch cuts, defined as non-Bloch FAs, which govern both GBZ geometries and OBC eigenvalue knots, as confirmed by our experiments. We finally demonstrate that geometry-induced non-Bloch EPs correspond to saddle points rather than branch points on the Riemann surface, with analogous geometries on GBZ surfaces. Our findings provide an alternative approach to realizing non-Bloch EPs and highlight the role of novel geometry features in non-Bloch band structures.

*Geometry-induced exceptional points*—To disentangle internal parameters such as hoppings and detunings with means unique in OBCs, we employ a multiband non-Hermitian system that hosts Bloch EPs and vary solely its geometry to drive the transition from Bloch to non-Bloch regimes. We thus utilize the reciprocal Lieb lattice shown at the top of Fig. 1(a). The Hamiltonian reads $H = \sum_{k}[i\gamma c^{\dagger}_{k,A}c_{k,A} - i\gamma c^{\dagger}_{k,C}c_{k,C} + (T_{AB} c^{\dagger}_{k,A}c_{k,B} + T_{AC}c^{\dagger}_{k,A}c_{k,C} + T_{BC} c^{\dagger}_{k,B}c_{k,C} + \text{h.c.})]$ [75], where $T_{AB} = 2t\cos(k_y/2)$, $T_{BC} = 2t\cos(k_x/2)$, and $T_{AC} = 4w\cos(k_x/2)\cos(k_y/2)$. Figure 1(b) depicts its experimental realization in an active acoustic lattice of coupled cavities, where the required hopping and gain/loss terms are precisely controlled by programmable microphones and speakers [76]. The Bloch band structure [Fig. 1(a), middle panel] consisting of three Re($E$) Riemann surfaces, with EPs identified by vanishing phase rigidity $r_j = \langle \psi^L_j | \psi^R_j \rangle / \langle \psi^R_j | \psi^R_j \rangle$, where $\psi^R$ ($\psi^L$) is the right (left) eigenstate of the $j$-th band [3,29,77,78]. Eight order-2 Bloch EPs are branch points connected by branch cuts (non-Hermitian FAs) [2,46-48], as sketched at the bottom of Fig. 1(a). These order-2 EPs are classified by braid words $\tau_n^{\pm 1}$ and their conjugate classes, with superscripts representing eigenvalue winding numbers and subscripts the band gaps [7-14].

We then compare the rectangular [Fig. 1(c)] and parallelogram [Fig. 1(d)] geometries to demonstrate Bloch and non-Bloch EPs. In the rectangular geometry, spectral reciprocity is preserved along both $k_x$ and $k_y$, eliminating skin modes. In contrast, the parallelogram geometry breaks reciprocity along $k_x$ but retains it along $k_b$, leading to geometry-dependent skin modes [58,79-82]. To experimentally locate EPs, we measure the system-wide Green's function by point-exciting individual sites and recording the complete multisite response, and its diagonalization yields the OBC spectrum and eigenstates [76].

Figures 1(c) and 1(d) show measured OBC spectra, color-coded by the phase rigidity of each eigenstate, whose minima indicate EPs. For the rectangular geometry [Fig. 1(c)], the observed EPs coincide with the Bloch EPs of Fig. 1(a), consistent with the absence of skin modes, as substantiated by the shown representative eigenstate and the identical PBC and OBC spectra. In contrast, the parallelogram geometry [Fig. 1(d)] exhibits a pronounced spectral deformation, signaling the onset of skin modes, as confirmed by the localized OBC eigenstate in the inset. The minima of phase rigidity reveal that the Bloch EPs ($\sim 1038.0 - 6.0i$ Hz, $1040.0 - 4.2i$ Hz, and $1040.0 - 7.8i$ Hz) vanish while non-Bloch EPs appear at a distinct energy ($\sim 1035.8 - 5.8i$ Hz). These emerged, geometry-induced EPs are thus non-Bloch EPs, demanding analysis from the non-Bloch perspective.

*Whitney cusps and non-Bloch Fermi arcs*—Since skin modes are present (absent) along the $x$ ($b$) direction, we utilize nonorthogonal $x$ and $b$ axes together with the primitive unit cell [dashed box in Fig. 2(a)], and extend $k_x$ as $\beta = e^{ik_x}$ while keeping $k_b \in [0, 2\pi]$. The corresponding non-Bloch Hamiltonian reads

$$h(\beta, k_b) = \begin{pmatrix} i\gamma & T_{AB}^- & T_{AC}^- \\ T_{AB}^+ & 0 & T_{BC}^- \\ T_{AC}^+ & T_{BC}^+ & -i\gamma \end{pmatrix}, \tag{1}$$

where $T_{AB}^+ = t(1 + \beta e^{-ik_b})$, $T_{BC}^+ = t(1 + \beta)$, $T_{AC}^+ = w(1 + \beta + \beta e^{ik_b} + \beta^2 e^{-ik_b})$, and $T^- = (T^+)^*$. Unlike the Bloch case, Eq. (1) indicates $h(\beta^{-1}, k_b) \neq h(\beta, k_b)$, implying the skin modes along the $x$ direction. Given such a complex continuation, the branch points of Riemann surfaces for $h(\beta, k_b)$ are singularities but do not necessarily reside on the OBC spectrum. We first locate all such non-Bloch branch points by the resultant $\text{Res}_E[f(\beta, E), \partial_E f(\beta, E)] = 0$, where $f(\beta, E) = \det[h(\beta) - EI]$ is the characteristic polynomial. Vanishing both real and imaginary parts of the resultant is required because the nonzero $w$ breaks non-Hermitian chiral symmetry $\Sigma^{-1} h^\dagger(\beta, k_b) \Sigma = -h(\beta, k_b)$ with $\Sigma = \text{diag}(1, -1, 1)$. Consequently, the non-Bloch branch points form continuous lines in the $(\beta, k_b)$-parameter space [83,84]. Figure 2(b) exhibits these branch-point lines, colored by braid words and oriented via the right-hand rule using their eigenvalue winding numbers. Bloch EPs undoubtedly reside on these lines, identical to the positions in Fig. 1(a). After identifying non-Bloch branch points, we next analyze GBZs to examine whether these manifest in the OBC spectrum.

GBZs predict OBC spectra by enforcing standing-wave conditions along the $x$ direction, yielding two-dimensional GBZ surfaces in the $(\beta, k_b)$-parameter space [49,62]. In multiband systems, each band typically has its own sub-GBZs, which become inherently inseparable in the presence of branch points. We acquire these sub-GBZs of Eq. (1) by vanishing $f(\beta, E)$ and the GBZ condition ($|\beta_2| = |\beta_3|$) [62,85,86]. To see whether non-Bloch branch points reside on the OBC spectrum, we find the intersection between GBZs and non-Bloch branch-point lines. Figure 2(c) shows two such sub-GBZ surfaces, GBZ$_1$ and GBZ$_2$, ordered by ascending $\text{Re}(E)$ Riemann surfaces. The surfaces exhibit two characteristic geometric features: Whitney cusps (stars) and a saddle point (diamond). The Whitney cusps anchor non-Bloch FAs and govern the topology of OBC eigenvalue knots, while the saddle point corresponds to the geometry-induced non-Bloch EP observed experimentally in Fig. 1(d). We detail them individually below.

The inset in Fig. 2(c) magnifies the GBZ surface near the blue star, revealing a geometric transition from a self-intersection to a smooth morphology—geometrically identified as a Whitney cusp (see Sec. I in Ref. [87]). Figure 3(a) plots GBZs for three representative $k_b$ values around the critical point $k_{WC} = 2.313$, together with the Re($E$) Riemann surface at $k_{WC}$. As $k_b$ decreases, GBZ$_1$ evolves from residing on a single Riemann surface to successively traversing two Riemann surfaces. When $k_b < k_{WC}$, GBZ$_1$ (darker blue) crosses the non-Bloch branch cut, forms a closed loop encircling the $\tau_1$ branch point on the adjacent Riemann surface, and returns to its original Riemann surface. The branch cuts connecting these non-Bloch branch points are exactly non-Bloch FAs, which correspond to contours of equal Re($E$) and are denoted as non-Bloch FA$_1$, with the subscript following the braid word of the linked branch points (see Sec. II in Ref. [87]). These GBZ and Riemann structures together indicate that in the nontrivial case, the branch cut is traversed twice while the branch point is encircled once. At $k_{WC}$, only GBZ$_1$ approaches the branch point, producing a sharp cusp rather than an intersection of two sub-GBZs. As exhibited in Fig. 3(c), no qualitative spectral transitions suggest that the blue star, although lying on the same branch-point line as the Bloch EP [blue circle in Fig. 2(b), $k_b \sim 4.212$], is not itself an EP. Hence, geometry-induced non-Bloch EPs observed in Fig. 1(d) are not trivial non-Bloch analogues of blue Bloch EPs ($1038.0 - 6.0i$ Hz) identified in Figs. 1(a) and 1(c), but rather distinct singularities arising from the GBZ geometry.

***Saddle point as geometry-induced exceptional point***—To clarify geometry-induced EPs, we examine another characteristic geometric feature, i.e., the saddle point at $k_b = k_{SP} = 1.878$ in Fig. 2(c). Figures 3(b) and 3(d) display the GBZs and corresponding spectra for three representative $k_b$ values around $k_{SP}$, together with the Re($E$) Riemann surface at $k_{SP}$ shown in Fig. 3(b). As $k_b$ decreases across $k_{SP}$, two GBZ branches coalesce and then separate within one Riemann surface, driving a spectral transition from a Re($E$) to an Im($E$) line gap [3,88,89]. Spectrally, this behavior resembles a parity-time phase transition, with a non-Bloch EP emerging at $k_{SP}$ (red diamond), coinciding with the experimentally observed minimum of phase rigidity around $E \sim 1035.8 - 5.8i$ Hz in Fig. 1(d). The gap-size scaling $|k_b - k_{SP}|^{1/2}$ further confirm its EP nature (see Sec. III in Ref. [87]). It clarifies geometry-induced EPs and, more significantly, implies that the GBZ geometry [Fig. 2(c)] acts as the base manifold and fundamentally alters the adhered Riemann surfaces.

To visualize, Fig. 3(e) plots Re($E$) Riemann surfaces as a function of the GBZ argument $k_x = \arg(\beta)$ and $k_b$ near $k_{WC}$ and $k_{SP}$. Unlike typical Bloch scenarios, Fig. 3(e) breaks the relation among branch cuts, branch points, and Bloch EPs shown in Fig. 1(a). The non-Bloch branch points are no longer EPs, while saddle points on the Riemann surfaces are non-Bloch EPs. It arises because each band has its own GBZ, serving as the base manifold. In the Bloch limit, this manifold is a single torus shared by all bands, but becomes multiple tori deformed along the $k_x$ direction for the parallelogram defined in Fig. 1(d). The right panel of Fig. 3(e) sketches two sub-GBZs glued together by non-Bloch FAs and Whitney cusps, while the geometry-induced non-Bloch EP resides on a single sub-GBZ. Such a detachment of EPs from FAs in non-Bloch band structures challenges conventional characterization schemes, as identifying EPs and branch points of FAs is the same in Bloch systems and is inherently rooted in eigenvalue braiding.

*Measured OBC eigenvalue knots*—To resolve them, we employ the OBC eigenvalue knot, which traces the evolution of OBC energies by varying $\arg(\beta)$. Figure 4 shows the OBC eigenvalue knots of five representative $k_b$ values (bottom panels), and the corresponding two sub-GBZs defining the parametric loops (top panels). When $k_b > k_{\text{WC}}$ [Fig. 4(a)], the two sub-GBZs are disconnected, and the resulting eigenvalue knots are likewise distinct. As $k_b$ decreases below $k_{\text{WC}}$ [Figs. 4(b) and 4(c)], GBZ$_1$ and GBZ$_2$ remain separate but each traverses a non-Bloch FA, leading to the OBC eigenvalue knots undergoing a type-I Reidemeister move [19], as shown in Fig. 4(c2). Comparing Figs. 4(a2) and 4(c2), we see that the total number of crossings in the knot diagram changes while all eigenvalue winding numbers remain zero, because the OBC spectrum does not enclose an area. This contrasts with Bloch EPs and reflects the correspondence that the superscript in braid words actually relates to the writhe Wr of eigenvalue knots rather than their winding numbers (see Sec. IV in Ref. [87]).

Upon further decreasing $k_b$ to $k_{\text{SP}}$, GBZ$_1$ and GBZ$_2$ intersect on the same Riemann surface, generating the geometry-induced non-Bloch EPs [red diamonds in Fig. 4(d)] and leading to OBC eigenvalue knots merging into ill-defined configurations [Fig. 4(d2)]. When $k_b < k_{\text{SP}}$, the non-Bloch EPs disappear, and two sub-GBZs become separated again, leaving the Im($E$) line gap in the knot diagram [Fig. 4(e)]. Therefore, measuring OBC eigenvalue knots provides an experimental route to identify non-Bloch FAs, branch points, and EPs.

We thus employ the non-Bloch supercell framework to probe non-Bloch band structures experimentally [90] (see Sec. V in Ref. [87]). By using the obtained data, we display the sub-GBZs and corresponding eigenvalue knots for three representative values of $k_b$, as shown by circles and spheres in Fig. 4. Figures 4(a) and 4(c) identify non-Bloch FAs experimentally, while Figs. 4(c) and 4(e) together confirm that geometry-induced EPs observed in Fig. 1(d) originate from the saddle points. These results firmly demonstrate that geometry-induced non-Bloch EPs are not located at the branch points of non-Bloch FAs.

*Conclusions*—In brief, we reveal that non-Bloch EPs are feasible to be induced by geometry and detach from non-Bloch FAs that remain determining OBC eigenvalue knots. Using a reciprocal non-Hermitian Lieb lattice as a prototype, we show that geometry-induced EPs emerge from deformations of GBZ surfaces under different geometric shapes. Distinct algebraic-geometric features are identified in both the non-Bloch band structure and GBZ surfaces, including Whitney cusps associated with non-Bloch branch points exactly on the GBZ and a saddle point corresponding to the non-Bloch EP. Branch points, even carrying OBC eigenenergies, are distinct from non-Bloch EPs, unlike their unified counterparts in Bloch band structures. By analyzing the transitions of OBC eigenvalue knots, we further find that non-Bloch branch points generate type-I Reidemeister moves in the knot diagram, whereas non-Bloch EPs drive the knot to undergo a transition in line-gap type. These findings underscore the critical role of base manifolds in non-Bloch band structures, giving rise to EPs and Whitney cusps unique to lattices with non-Hermitian skin effect. Geometry-induced EPs thus introduce an additional geometric control parameter for tailoring EP-based functionalities, enabling non-Hermitian devices whose topological and spectral features can be reconfigured purely by mechanical deformation or boundary shaping, without modifying material composition.

**Acknowledgements**

We thank Prof. Guancong Ma for the helpful discussions. This work is supported by the National Key R&D Program of China (No. 2022YFA1404500, No. 2022YFA1404701), the National Natural Science Foundation of China (No.12347144), and the Shanghai Science and Technology Innovation Action Plan (No. 24Z510205936). Y. J. thanks the support of startup funds from Penn State University and NSF awards 2039463 and 195122.

**References**

[1] Y. Ashida, Z. Gong, and M. Ueda, Adv. Phys. **69**, 249-435 (2020).

[2] E. J. Bergholtz, J. C. Budich, and F. K. Kunst, Rev. Mod. Phys. **93**, 015005 (2021).

[3] K. Ding, C. Fang, and G. Ma, Nat. Rev. Phys. **4**, 745-760 (2022).

[4] N. Okuma and M. Sato, Annu. Rev. Condens. Matter Phys. **14**, 83-107 (2023).

[5] Q. Zhong, M. Khajavikhan, D. N. Christodoulides, and R. El-Ganainy, Nat. Commun. **9**, 4808 (2018).

[6] C. Chen, L. Jin, and R.-B. Liu, New J. Phys. **21**, 083002 (2019).

[7] C. C. Wojcik, X.-Q. Sun, T. Bzdušek, and S. Fan, Phys. Rev. B **101**, 205417 (2020).

[8] H. Hu and E. Zhao, Phys. Rev. Lett. **126**, 010401 (2021).

[9] K. Wang, A. Dutt, K. Y. Yang, C. C. Wojcik, J. Vučković, and S. Fan, Science **371**, 1240-1245 (2021).

[10] K. Wang, A. Dutt, C. C. Wojcik, and S. Fan, Nature (London) **598**, 59-64 (2021).

[11] H. Hu, S. Sun, and S. Chen, Phys. Rev. Res. **4**, L022064 (2022).

[12] Y. S. S. Patil, J. Höller, P. A. Henry, C. Guria, Y. Zhang, L. Jiang, N. Kralj, N. Read, and J. G. E. Harris, Nature (London) **607**, 271-275 (2022).

[13] C. C. Wojcik, K. Wang, A. Dutt, J. Zhong, and S. Fan, Phys. Rev. B **106**, L161401 (2022).

[14] C.-X. Guo, S. Chen, K. Ding, and H. Hu, Phys. Rev. Lett. **130**, 157201 (2023).

[15] W. B. Rui, Y. X. Zhao, and Z. D. Wang, Phys. Rev. B **108**, 165105 (2023).

[16] J. Hu, R.-Y. Zhang, Y. Wang, X. Ouyang, Y. Zhu, H. Jia, and C. T. Chan, Nat. Phys. **19**, 1098-1103 (2023).

[17] C. Lv and Q. Zhou, Phys. Rev. D **110**, 084039 (2024).

[18] H. Jia, J. Hu, R.-Y. Zhang, Y. Xiao, D. Wang, M. Wang, S. Ma, X. Ouyang, Y. Zhu, and C. T. Chan, Phys. Rev. Lett. **134**, 206603 (2025).

[19] Z. Li, K. Ding, and G. Ma, Phys. Rev. Res. **5**, 023038 (2023).

[20] Q. Zhang, Y. Li, H. Sun, X. Liu, L. Zhao, X. Feng, X. Fan, and C. Qiu, Phys. Rev. Lett. **130**, 017201 (2023).

[21] Q. Zhang, L. Zhao, X. Liu, X. Feng, L. Xiong, W. Wu, and C. Qiu, Phys. Rev. Res. **5**, L022050 (2023).

[22] Y. Long, Z. Wang, C. Zhang, H. Xue, Y. X. Zhao, and B. Zhang, Phys. Rev. Lett. **132**, 236401 (2024).

[23] Z. Rao, C. Meng, Y. Han, L. Zhu, K. Ding, and Z. An, Nat. Phys. **20**, 1904-1911 (2024).

[24] H. Zhang, T. Liu, Z. Xiang, K. Xu, H. Fan, and D. Zheng, PRX Quantum **6**, 020328 (2025).

[25] C. Dembowski, H. D. Gräf, H. L. Harney, A. Heine, W. D. Heiss, H. Rehfeld, and A. Richter, Phys. Rev. Lett. **86**, 787-790 (2001).

[26] S.-B. Lee, J. Yang, S. Moon, S.-Y. Lee, J.-B. Shim, S. W. Kim, J.-H. Lee, and K. An, Phys. Rev. Lett. **103**, 134101 (2009).

[27] B. Dietz, H. L. Harney, O. N. Kirillov, M. Miski-Oglu, A. Richter, and F. Schäfer, Phys. Rev. Lett. **106**, 150403 (2011).

[28] T. Gao, E. Estrecho, K. Y. Bliokh, T. C. H. Liew, M. D. Fraser, S. Brodbeck, M. Kamp, C. Schneider, S. Höfling, Y. Yamamoto, F. Nori, Y. S. Kivshar, A. G. Truscott, R. G. Dall, and E. A. Ostrovskaya, Nature (London) **526**, 554-558 (2015).

[29] K. Ding, G. Ma, M. Xiao, Z. Q. Zhang, and C. T. Chan, Phys. Rev. X **6**, 021007 (2016).

[30] C. Hahn, Y. Choi, J. W. Yoon, S. H. Song, C. H. Oh, and P. Berini, Nat. Commun. **7**, 12201 (2016).

[31] D. Zhang, X.-Q. Luo, Y.-P. Wang, T.-F. Li, and J. Q. You, Nat. Commun. **8**, 1368 (2017).

[32] K. Ding, G. Ma, Z. Q. Zhang, and C. T. Chan, Phys. Rev. Lett. **121**, 085702 (2018).

[33] L. Xiao, T. Deng, K. Wang, Z. Wang, W. Yi, and P. Xue, Phys. Rev. Lett. **126**, 230402 (2021).

[34] J. Wiersig, Phys. Rev. Lett. **112**, 203901 (2014).

[35] W. Chen, Ş. Kaya Özdemir, G. Zhao, J. Wiersig, and L. Yang, Nature (London) **548**, 192-196 (2017).

[36] H. Hodaei, A. U. Hassan, S. Wittek, H. Garcia-Gracia, R. El-Ganainy, D. N. Christodoulides, and M. Khajavikhan, Nature (London) **548**, 187-191 (2017).

[37] Y.-H. Lai, Y.-K. Lu, M.-G. Suh, Z. Yuan, and K. Vahala, Nature (London) **576**, 65-69 (2019).

[38] K. Bai, L. Fang, T.-R. Liu, J.-Z. Li, D. Wan, and M. Xiao, Natl. Sci. Rev. **10**, nwac259 (2022).

[39] W. Mao, Z. Fu, Y. Li, F. Li, and L. Yang, Sci. Adv. **10**, eadl5037 (2024).

[40] B. Peng, Ş. K. Özdemir, M. Liertzer, W. Chen, J. Kramer, H. Yılmaz, J. Wiersig, S. Rotter, and L. Yang, Proc. Natl. Acad. Sci. U.S.A. **113**, 6845-6850 (2016).

[41] L. Jin and Z. Song, Phys. Rev. Lett. **121**, 073901 (2018).

[42] J. Zhang, B. Peng, Ş. K. Özdemir, K. Pichler, D. O. Krimer, G. Zhao, F. Nori, Y.-x. Liu, S. Rotter, and L. Yang, Nat. Photonics **12**, 479-484 (2018).

[43] J. Doppler, A. A. Mailybaev, J. Böhm, U. Kuhl, A. Girschik, F. Libisch, T. J. Milburn, P. Rabl, N. Moiseyev, and S. Rotter, Nature (London) **537**, 76-79 (2016).

[44] H. Xu, D. Mason, L. Jiang, and J. G. E. Harris, Nature (London) **537**, 80-83 (2016).

[45] F. K. Kunst and V. Dwivedi, Phys. Rev. B **99**, 245116 (2019).

[46] H. Zhou, C. Peng, Y. Yoon, C. W. Hsu, K. A. Nelson, L. Fu, J. D. Joannopoulos, M. Soljačić, and B. Zhen, Science **359**, 1009-1012 (2018).

[47] V. Kozii and L. Fu, Phys. Rev. B **109**, 235139 (2024).

[48] R. Zheng, J. Lin, J. Liang, K. Ding, J. Lu, W. Deng, M. Ke, X. Huang, and Z. Liu, Commun. Phys. **7**, 298 (2024).

[49] S. Yao and Z. Wang, Phys. Rev. Lett. **121**, 086803 (2018).

[50] F. Song, S. Yao, and Z. Wang, Phys. Rev. Lett. **123**, 170401 (2019).

[51] N. Okuma, K. Kawabata, K. Shiozaki, and M. Sato, Phys. Rev. Lett. **124**, 086801 (2020).

[52] L. Li, C. H. Lee, S. Mu, and J. Gong, Nat. Commun. **11**, 5491 (2020).

[53] S. Weidemann, M. Kremer, T. Helbig, T. Hofmann, A. Stegmaier, M. Greiter, R. Thomale, and A. Szameit, Science **368**, 311-314 (2020).

[54] T. Helbig, T. Hofmann, S. Imhof, M. Abdelghany, T. Kiessling, L. W. Molenkamp, C. H. Lee, A. Szameit, M. Greiter, and R. Thomale, Nat. Phys. **16**, 747-750 (2020).

[55] L. Xiao, T. Deng, K. Wang, G. Zhu, Z. Wang, W. Yi, and P. Xue, Nat. Phys. **16**, 761-766 (2020).

[56] A. Ghatak, M. Brandenbourger, J. van Wezel, and C. Coulais, Proc. Natl. Acad. Sci. U.S.A. **117**, 29561-29568 (2020).

[57] X. Zhang, Y. Tian, J.-H. Jiang, M.-H. Lu, and Y.-F. Chen, Nat. Commun. **12**, 5377 (2021).

[58] K. Zhang, Z. Yang, and C. Fang, Nat. Commun. **13**, 2496 (2022).

[59] Y. O. Nakai, N. Okuma, D. Nakamura, K. Shimomura, and M. Sato, Phys. Rev. B **109**, 144203 (2024).

[60] B.-B. Wang, Z. Cheng, H.-Y. Zou, Y. Ge, K.-Q. Zhao, Q.-R. Si, S.-Q. Yuan, H.-X. Sun, H. Xue, and B. Zhang, Proc. Natl. Acad. Sci. U.S.A. **122**, e2422154122 (2025).

[61] K. Zhang, C. Shu, and K. Sun, Phys. Rev. X **15**, 031039 (2025).

[62] K. Yokomizo and S. Murakami, Phys. Rev. Lett. **123**, 066404 (2019).

[63] Y. Li, X. Ji, Y. Chen, X. Yan, and X. Yang, Phys. Rev. B **106**, 195425 (2022).

[64] Y. Fu and Y. Zhang, Phys. Rev. B **110**, L121401 (2024).

[65] K. Wang, T. Li, L. Xiao, Y. Han, W. Yi, and P. Xue, Phys. Rev. Lett. **127**, 270602 (2021).

[66] Z. Lei, C. H. Lee, and L. Li, Commun. Phys. **7**, 100 (2024).

[67] M. Xu, Z. Gong, and W. Yi, Phys. Rev. B **111**, 214305 (2025).

[68] K. Yokomizo and S. Murakami, Phys. Rev. Res. **2**, 043045 (2020).

[69] Y. Fu and S. Wan, Phys. Rev. B **105**, 075420 (2022).

[70] Y. Fu and Y. Zhang, Phys. Rev. B **107**, 115412 (2023).

[71] Y.-M. Hu, H.-Y. Wang, Z. Wang, and F. Song, Phys. Rev. Lett. **132**, 050402 (2024).

[72] B. Teissier, *Varietes polaires II Multiplicites polaires, sections planes, et conditions de whitney* (Springer, Berlin, 1982).

[73] M. Helmer, G. Papathanasiou, and F. Tellander, arXiv:2402.14787.

[74] M. Helmer and R. Mohr, arXiv:2406.17122.

[75] Y.-X. Xiao, K. Ding, R.-Y. Zhang, Z. H. Hang, and C. T. Chan, Phys. Rev. B **102**, 245144 (2020).

[76] J.-X. Zhong, J. Kim, K. Chen, J. Lu, K. Ding, and Y. Jing, Phys. Rev. B **112**, L220301 (2025).

[77] I. Rotter, J. Phys. A: Math. Theor. **42**, 153001 (2009).

[78] W. Tang, X. Jiang, K. Ding, Y.-X. Xiao, Z.-Q. Zhang, C. T. Chan, and G. Ma, Science **370**, 1077-1080 (2020).

[79] Q. Zhou, J. Wu, Z. Pu, J. Lu, X. Huang, W. Deng, M. Ke, and Z. Liu, Nat. Commun. **14**, 4569 (2023).

[80] T. Wan, K. Zhang, J. Li, Z. Yang, and Z. Yang, Sci. Bull. **68**, 2330-2335 (2023).

[81] W. Wang, M. Hu, X. Wang, G. Ma, and K. Ding, Phys. Rev. Lett. **131**, 207201 (2023).

[82] K. Zhang, Z. Yang, and K. Sun, Phys. Rev. B **109**, 165127 (2024).

[83] K. Kawabata, T. Bessho, and M. Sato, Phys. Rev. Lett. **123**, 066405 (2019).

[84] S. Sayyad, M. Stålhammar, L. Rødland, and F. K. Kunst, Scipost Phys. **15**, 200 (2023).

[85] Z. Yang, K. Zhang, C. Fang, and J. Hu, Phys. Rev. Lett. **125**, 226402 (2020).

[86] Y.-M. Hu and Z. Wang, Phys. Rev. Res. **5**, 043073 (2023).

[87] See Supplemental Material for details on (I) Whitney cusp and Whitney stratification; (II) details of Whitney cusps and non-Bloch Fermi arcs; (III) asymptotic analysis of gap sizes near geometry-induced EPs; (IV) knot transition near another non-Bloch EP; (V) non-Bloch supercell framework for experimental measurement.

[88] K. Kawabata, K. Shiozaki, M. Ueda, and M. Sato, Phys. Rev. X **9**, 041015 (2019).

[89] S. Verma and M. J. Park, Commun. Phys. **7**, 21 (2024).

[90] J.-X. Zhong, J. Lin, K. Chen, J. Lu, K. Ding, and Y. Jing, arXiv:2510.20160.

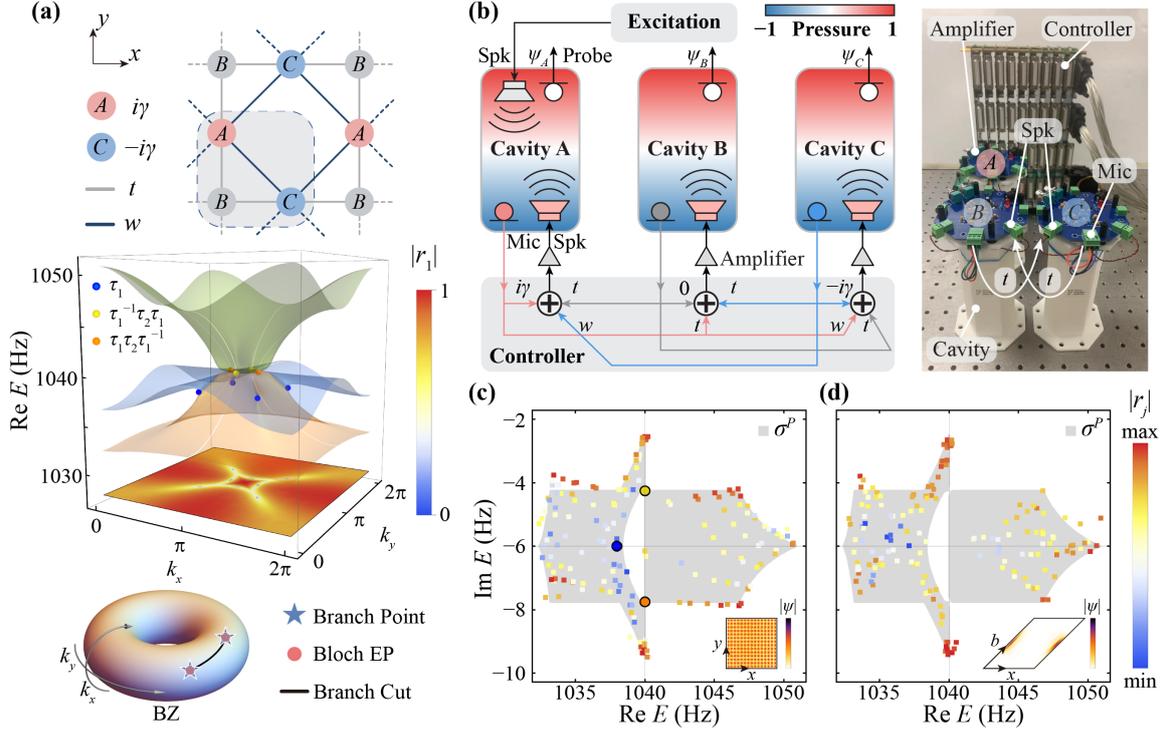

FIG. 1. (a) Schematic of the non-Hermitian Lieb lattice (top). $A$ ($C$) sites exhibit onsite gain $i\gamma$ (loss $-i\gamma$), while the nearest-neighbor (next-nearest-neighbor) hoppings are $t$ ($w$). The middle and bottom panels show one representative $\text{Re}(E)$ Riemann surface under PBCs and the BZ torus, respectively. The spheres on the Riemann surface represent order-2 EPs with their braid words labeled, and the lower plane depicts the phase rigidity of the orange band. Blue stars and spheres denote branch points and Bloch EPs, connected by a branch cut (black line). (b) Experimental realization of the designed lattice. A schematic (photograph) of the unit cell is shown on the left (right). Each acoustic cavity is equipped with microphones (Mic) and loudspeakers (Spk) for the real-time controller to implement hoppings and onsite potentials. Another set of microphones and loudspeakers is utilized to perform the pump-probe measurement of the system-wise Green's function. (c,d) Experimental OBC spectra for a rectangle (c) and a parallelogram with one internal angle of $\pi/4$ (d). The color of each OBC eigenfrequency encodes the phase rigidity of measured eigenstates. Insets show respective geometries alongside a representative OBC eigenstate. The number of unit cells along each edge direction is denoted by $N_x$, $N_y$, and $N_b$. The gray region indicates the PBC spectrum range $\sigma^P$. Blue, yellow, and orange circles correspond to Bloch EPs identified in (a). The experimental parameters are $\omega_0 = 1040 - 6.0i$ Hz, $t = \gamma = 3.5$ Hz, $w = 0.7$ Hz, and $N_y = N_b = 6$. Two experimental sets of OBC spectra, $N_x = 4$ and $N_x = 5$, are shown superimposed in (c) and (d).

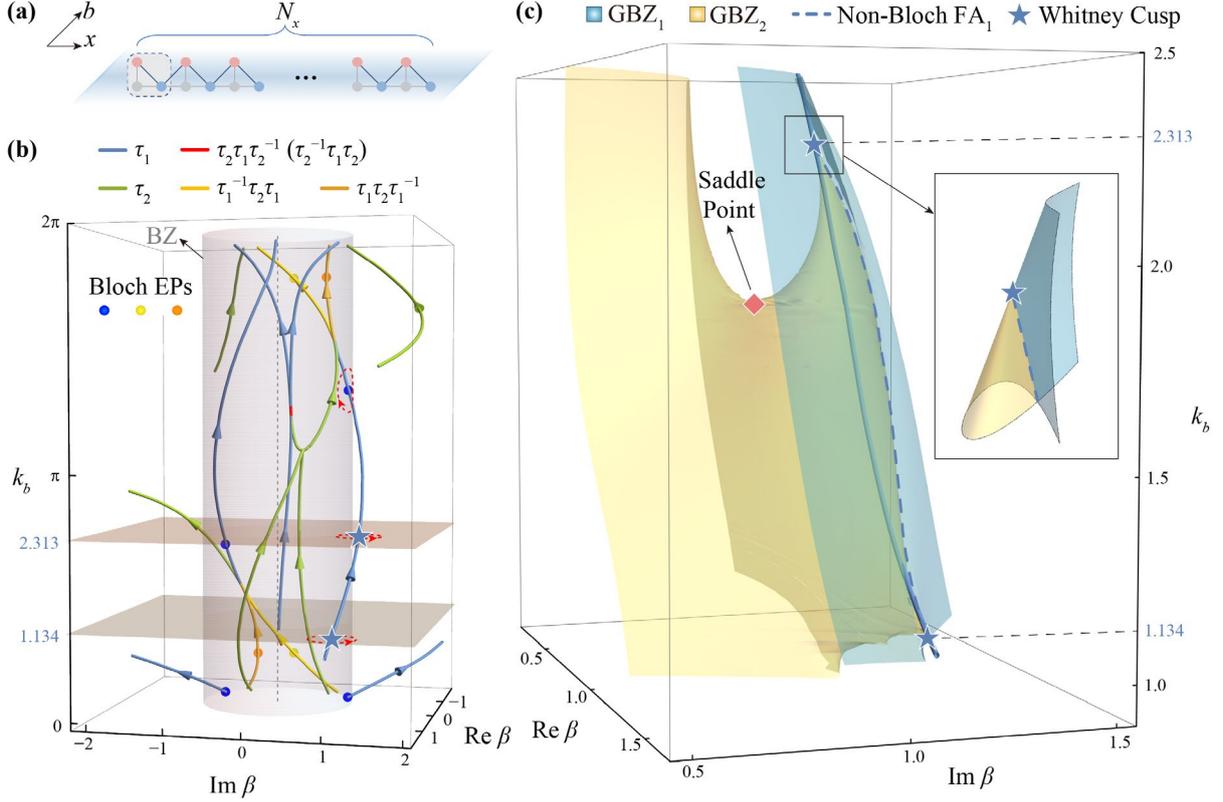

FIG. 2. (a) Quasi-one-dimensional lattice with PBCs (OBCs) applied along the $b$ direction ($x$ direction). (b) Continuous lines of non-Bloch branch points in the complex $\beta$ and $k_b$ space, with colors representing different braid words. The gray cylinder denotes the BZ ($|\beta| = 1$) containing eight Bloch EPs. Blue stars on the $k_b = 1.134$ and $k_b = 2.313$ planes mark the intersection between non-Bloch branch points and the GBZ. (c) Two sub-GBZs, GBZ$_1$ (blue) and GBZ$_2$ (yellow), together with the branch-point line (solid blue) spanning the $k_b$ range covering the two blue stars in (b). The dashed blue line denotes non-Bloch FA$_1$. The red diamond marks a saddle point on the GBZ$_2$ surface, and the inset magnifies the boxed region, showing a typical Whitney cusp (blue star) where the self-intersecting portion of the surface terminates.

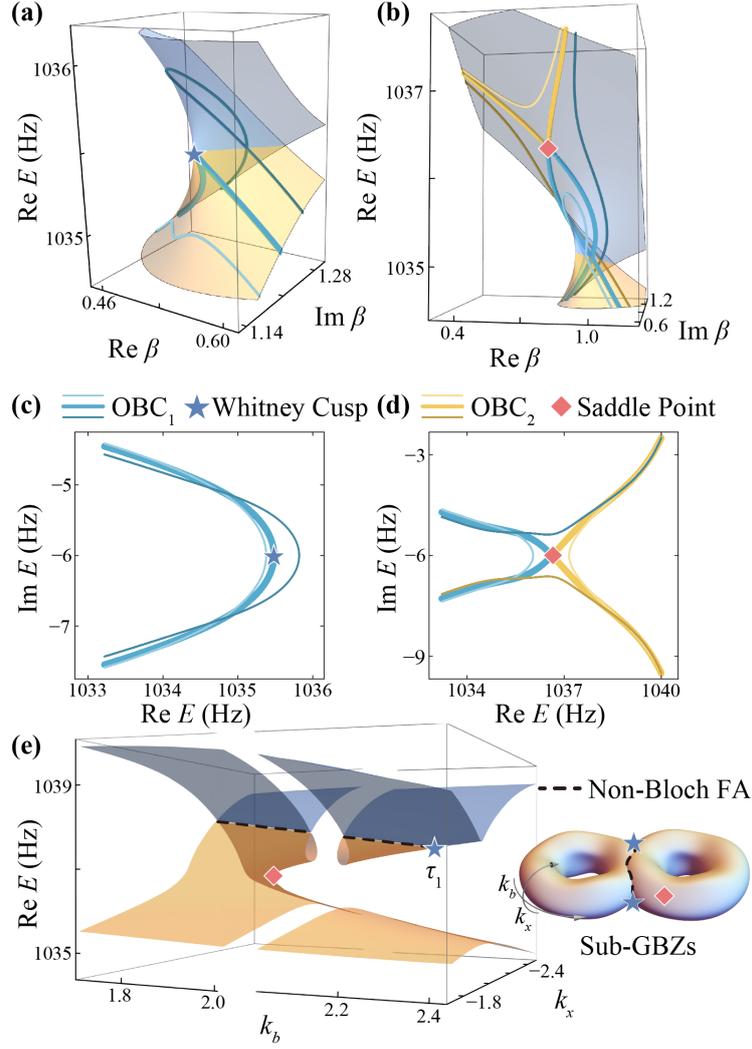

FIG. 3. (a,b) Evolution of the sub-GBZs (solid lines) as $k_b$ sweeps across the Whitney cusp at $k_{WC} = 2.313$ (a) and the saddle point at $k_{SP} = 1.878$ (b). The surfaces depict the Re($E$) Riemann surfaces at these two critical points. (c,d) Corresponding OBC spectra associated with the sub-GBZs in (a,b). Blue (yellow) lines in (a-d) represent GBZ$_1$ and OBC$_1$ (GBZ$_2$ and OBC$_2$), and lighter colors indicate larger $k_b$. (e) Non-Bloch band structure, Re($E$) as functions of $k_x$ and $k_b$, in the vicinity of the Whitney cusp and saddle point. The right-hand inset sketches the two sub-GBZs as deformed tori, highlighting the Whitney cusp (blue star), saddle point (red diamond), and non-Bloch FAs (black dashed lines).

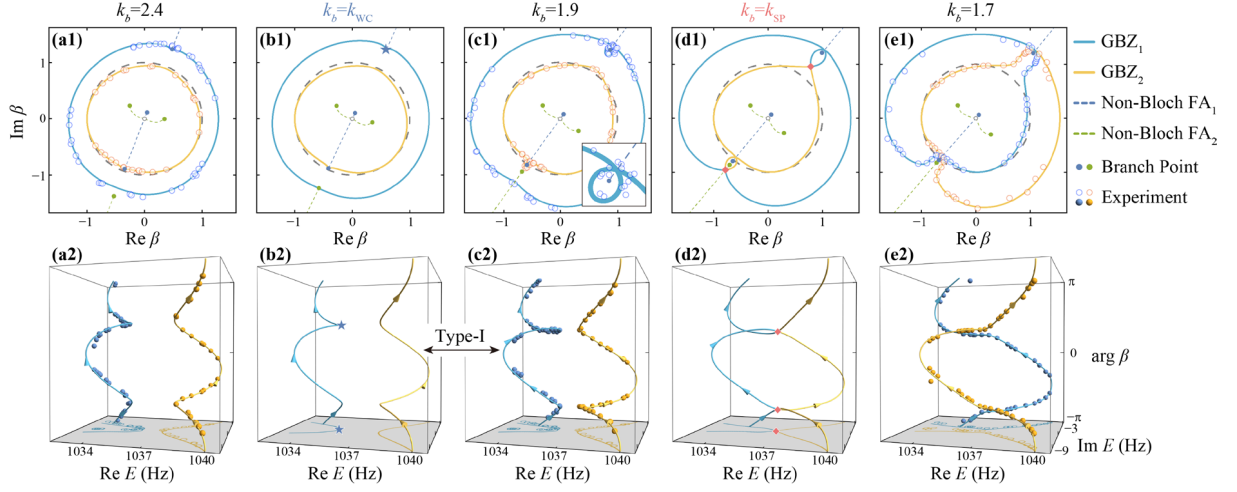

FIG. 4. Evolution of sub-GBZs (a1-e1) and OBC eigenvalue knot diagrams (a2-e2) for $k_b = 2.4$ (a), $k_{WC}$ (b), 1.9 (c), $k_{SP}$ (d), and 1.7 (e). Open circles (a1-e1) and spheres (a2-e2) denote measured results, while solid lines are theoretical ones. In (a1-e1), filled circles and dashed lines mark the non-Bloch branch points and branch cuts identified in Fig. 2(a). The blue stars in (b) and red diamonds in (d) indicate the Whitney cusp ($k_{WC} = 2.262$, $E_{WC} = 1035.4 - 6.0i$ Hz) and saddle point ($k_{SP} = 1.768$, $E_{SP} = 1036.6 - 6.0i$ Hz), respectively. Inset of (c1) magnifies the segment of GBZ$_1$ traversing non-Bloch FA$_1$, during which the eigenvalue knots undergo a type-I Reidemeister move as shown in (c2). All parameters in the non-Bloch supercell measurement are identical to those used for the OBC measurement in Fig. 1, except $w = 0.66$ Hz and $N_x = 15$.